\documentclass[10pt,twocolumn,letterpaper]{article}

\usepackage{iccv}
\usepackage{times}
\usepackage{epsfig}
\usepackage{graphicx}
\usepackage{amsmath}
\usepackage{amssymb}
\usepackage{enumitem}

\usepackage[pagebackref=true,breaklinks=true,letterpaper=true,colorlinks,bookmarks=false]{hyperref}

\iccvfinalcopy % *** Uncomment this line for the final submission

\def\iccvPaperID{10840} % *** Enter the ICCV Paper ID here

\ificcvfinal\fi

\author{{Hadeel Mabrouk,
Omar Abugabal,
Nourhan Sakr,
and Hesham M. Eraqi}\\
{\small \it Computer Science and Engineering Department, The American University in Cairo}\\
{\small \{hadeelmabrouk, abugabal, nouri.sakr, heraqi\}@aucegypt.edu}
}

\begin{document}

%%%%%%%%% TITLE Mixing Senses to Understand Lips
\title{Lip-Listening: Mixing Senses to Understand Lips using Cross Modality Knowledge Distillation for Word-Based Models}

\maketitle

% H. Eraqi proposed abstract edits (can be used as a hint that might improve the current abstract):
% Impressive progress in the domain of speech recognition has been exhibited by audio and audio-visual systems. Nevertheless, there is still much to be explored with regards to visual speech recognition systems due to the visual ambiguity of some phonemes. In this work, we propose adopting sequence-level and frame-level Knowledge Distillation (KD) from a speech recognizer to improve visual speech recognition. The proposed model follows an encoder-decoder architecture and leverages knowledge in audio data for training the visual speech recognition model. A Gaussian-shaped averaging alignment method is adopted in the frame-level KD to aid the model in distilling knowledge at the encoder level. The proposed model improves lip-reading state-of-the-art by 0.8\% on LRW benchmark.

% Impressive progress in the domain of speech recognition has been exhibited by audio and audio-visual systems. However, there is still much to be explored with regards to visual speech recognition systems due to the visual ambiguity of some phonemes. Furthermore, there currently exists a drastic discrepancy between the respective performances of audio and visual speech recognition models. Although existing audio models might achieve relatively higher performances in ideal  conditions, their performance tends to degrade significantly when presented with distorted or corrupted audio signals. 

%%%%%%%%% ABSTRACT
\begin{abstract}
   In this work, we propose a technique to transfer speech recognition capabilities from audio speech recognition systems to visual speech recognizers, where our goal is to utilize audio data during lipreading model training. Impressive progress in the domain of speech recognition has been exhibited by audio and audio-visual systems. Nevertheless, there is still much to be explored with regards to visual speech recognition systems due to the visual ambiguity of some phonemes. To this end, the development of visual speech recognition models is crucial given the instability of audio models. The main contributions of this work are i) building on recent state-of-the-art word-based lipreading models by integrating sequence-level and frame-level Knowledge Distillation (KD) to their systems; ii) leveraging audio data during training visual models, a feat which has not been utilized in prior word-based work; iii) proposing the Gaussian-shaped averaging in frame-level KD, as an efficient technique that aids the model in distilling knowledge at the sequence model encoder. This work proposes a novel and competitive architecture for lip-reading, as we demonstrate a noticeable improvement in performance, setting a new benchmark equals to 88.64\% on the LRW dataset.
\end{abstract}
% H. Eraqi: You can consider startin with a quick sentence introducing the problem and the importance to solve it (motive the reader).
% H. Eraqi: The abstract should better mention in the incremental improvement over the state-of-the-art results on our benchmark dataset, this hugely increases the acceptance chances. Also, it's nice to tell the improvement that KD adds.

%%%%%%%%% BODY TEXT
\section{Introduction}
Interest in the domain of speech recognition has exponentially risen since its emergence around the 1950s. Systems evolved from very basic models that could understand only digits, to modern day systems that we heavily rely on, e.g. digital assistants, real-time language translation, and closed captioning \cite{intro1}. While these contributions assisted in various daily tasks, there still remains a plethora of applications and models to explore within this domain. 

There are currently ongoing efforts to implement efficient models that can recognize speech using audio or audio-visual based systems \cite{paper2,Xu_2020_WACV,c6}, but very few that rely solely on visual input. Audio and audio-visual speech recognition models significantly outperform  visual models in terms of accuracy due to the following reasons. Given the multifaceted nature of video data, the challenging process of extracting the necessary visual features may significantly hinder the accuracy of a model, which may be decreased by including, for example, the angle of the recorded speaker and the lighting. Moreover, the inherent ambiguity that exists within language complicates the task even further. 

Speech is broken down into two atomic units of sound: phonemes and visemes. A phoneme represents the basic sound that a given letter produces when spoken, while a viseme is the visual equivalent of that phoneme. Yet, there is no direct mapping between phonemes and visemes. Besides, two or more distinct phonemes may share the same viseme, such as guttural sounds like a `p' and a `b', which have nearly indistinguishable lip movements but produce distinct sounds. As such, these types of recognition tasks require the context of a given letter or word with respect to neighboring letters or words.

While audio and audio-visual models are generally of higher accuracy than visual models, they still have faults. Yet, models that rely heavily on audio are subject to drastic declines in performance metrics, e.g. when audio signals are corrupted or distorted, or if the surrounding environment surpasses a certain noise threshold \cite{intro2}. This gap cements the importance of visual speech recognition, which is more complex than audio speech recognition. If these models continue to develop at the current rate, the near future may experience the development of new applications that efficiently detect and recognize speech regardless of how noisy the background environment is, or whether the model lacks audio input. Such developments could tremendously improve the quality of life of those with hearing impairments or  speech impediments by providing a new means of communication with the world \cite{intro2}. 

Within this framework, lip reading models started gaining their significance and recent work has achieved increasingly better accuracy due to advancements in deep learning techniques, as well as the emergence of large-scaled, publicly available lip reading data sets \cite{withoutpain,c10,c17,c16}.

Our work builds effectively on the model presented by \cite{withoutpain} to enhance the performance of lipreading systems. We focus on word-based models and provide experimental evidence demonstrating how the proposed model achieves new state-of-the-art results on the LRW dataset. Our main contributions are threefold:
\begin{itemize}[noitemsep]
    \item We incorporate a variant of multi-granularity knowledge distillation in the backend of the proposed lipreading model at multiple temporal scales, namely frame-level and sequence-level.
    \item We utilize both audio and visual aspects of the input videos in order to make use of the available data in its entirety. To our best knowledge, we are the first to do so for word-based visual speech recognizers. 
    \item We introduce the Gaussian-shaped averaging technique at the encoder-level cross modality distillation, as an efficient approach that leverages the temporal correlation between audio and visual features.
\end{itemize}

%------------------------------------------------------------------------
\section{Related work}
\paragraph{Traditional models vs. DNN models.} Prior to the development of Deep Neural Networks (DNNs), speech recognition architectures could be divided into three phases: lip localization, feature extraction and classification. Lip localization usually relied on Deformable Part Models (DPMs) \cite{localization1}, or Regression-based models \cite{localization2} for feature detection. Feature extraction relied either on Active Appearance Models (AAMs), Discrete Cosine Transforms (DCT), or a combination of DCT with either Linear Discriminant Analysis or Principal Component Analysis. Finally,  classification relied heavily on the use of Hidden Markov Models (HMMs) for the recognition of a given digit or character \cite{intro2}. 

As DNNs progressively became more sophisticated, traditional networks became increasingly obsolete. DNNs differ in the newfound reliance on CNNs for feature extraction \cite{c12,c13,c11} and the replacement of HMMs with recurrent networks, such as Long-Short Term Memory (LSTMs) \cite{c10,c9,c14}. Consequently, the recent development of speech recognition models shifted to entirely relying on deep learning, known as end-to-end DNN architectures. Researchers found that while DNN and traditional architectures perform similarly given simple tasks like letters or digit recognition, DNNs perform significantly better in more complex recognition tasks, by a margin of around 40\% when it comes to word-based models \cite{intro2}, which is the focus of our work.

\paragraph{Speech-producing vs text-producing models.} Past works were predominantly text-based \cite{c6,c1,mg,df}, with only a few instances of speech-based models \cite{c2} most likely since text-based models are relatively easier to implement. However, Text-to-Speech (TTS) technologies have advanced enough to make the task of converting text output to speech output relatively simple. Yet , there is no indication of one approach being superior to the other. It is usually a matter of which output is deemed more appropriate for the objectives of a given model. 

\begin{table}
\begin{center}
\begin{tabular}{|l|c|}
\hline
Model & Top-1 Acc. (\%) \\
\hline\hline
Multi-Grained \cite{mg} & 83.3 \\
Policy Gradient \cite{pg} & 83.5 \\
STFM Convolutional Sequence \cite{Zhang_2019_ICCV} & 83.7 \\
Deformation Flow \cite{df} & 84.1 \\
Two Stream \cite{ts} & 84.1 \\
Mutual Information \cite{mi} & 84.4 \\
Face Cutout \cite{c5} & 85.0 \\
Temporal Convolution \cite{c4} & 85.3 \\
Hierarchical Pyramidal Convolution \cite{chen2020lipreading} & 86.8 \\
Effective Lipreading without Pain \cite{withoutpain} & 88.4 \\
Multi-Stage Distillation \cite{msd} & 88.5 \\
\textbf{Ours} & \textbf{88.6} \\
\hline
\end{tabular}
\end{center}
\caption{Comparison between existing models in terms of Top-1 Accuracy vs. our proposed model}
\label{lit}
\end{table}

\begin{figure*}
\begin{center}
\includegraphics[width=6.5in]{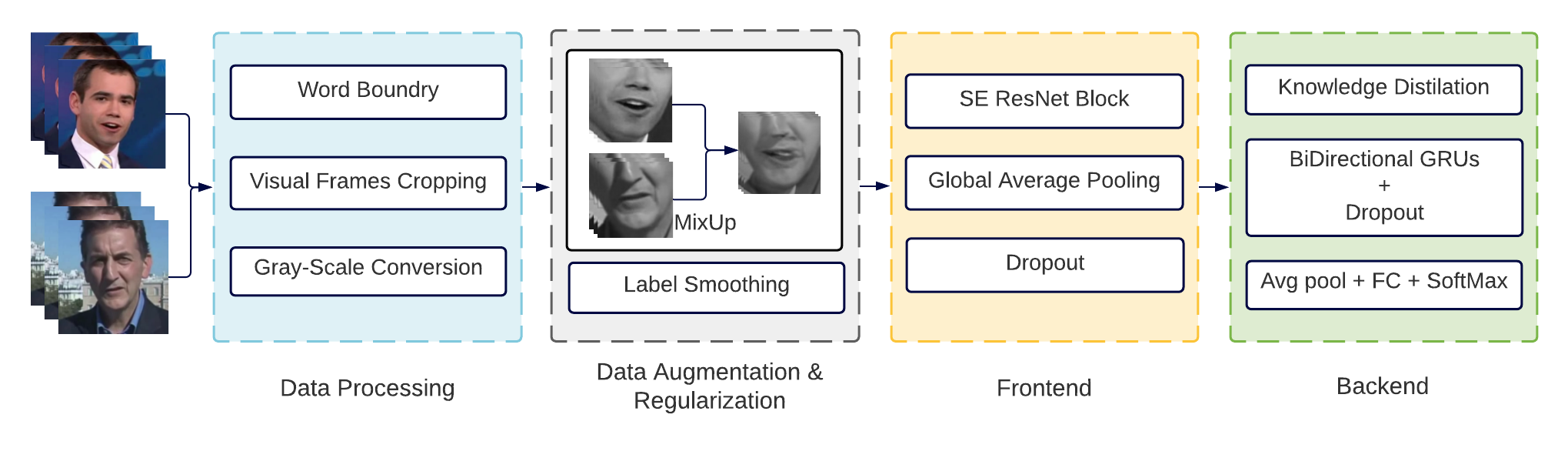}
\end{center}
   \caption{Our Proposed Pipeline}
\label{pipeline}
\end{figure*}

\paragraph{Word-based vs sentence-based models.} Lip-reading models can also be differentiated by the output they produce, whether it is a word or a series of characters forming a phrase or a sentence. Word-based models focus on words in isolation, while sentence-based models \cite{ma2021endtoend} can take into account the context in which this word is uttered. Sentence based models can also utilize the use of external language models to further leverage text-only data like what was proposed in  \cite{c6,c1}. With respect to word-based models, recent research in the area of visual speech recognition has resulted in a variety of models with different frontend and backend architectures. In such models, the feature extraction layers vary between ResNet-34 and ResNet-18 like in \cite{mg,ts,pg,mi}. On the other hand, the back-end layers vary between LSTM \cite{mg,ts}, BiGRU \cite{df,c5}, and recently MS-TCN \cite{msd, c4}, where knowledge distillation is introduced to offer light-weight lipreading models pursuing an effective deployment of lipreading in practical scenarios. Table \ref{lit} summarizes the results of Top-1 accuracy scores achieved by recent state-of-the-art models in the word-based lipreading domain.

\paragraph{Cross modality knowledge distillation.} Proven to be successful in object detection among other computer vision tasks, Knowledge Distillation (KD) has also been used in automatic lipreading \cite{Deng_2019_ICCV,Hou_2019_ICCV,c14}. Since audio recognition systems currently tend to be a quite mature problem achieving relatively higher accuracy, visual speech recognition models are still relatively under-explored. Thus, by distilling the knowledge transferred from the audio model to the visual model, \cite{c14,c8} propose that it improves the accuracy of sentence-based visual speech recognition systems. Additionally, \cite{c8} show that this approach can also be used to leverage arbitrary amounts of unlabeled video data to improve the model's performance.

In \cite{c14}, the authors apply knowledge distillation on LRS2 and CLMR dataset by introducing a new loss function, that in addition to the base cross entropy loss, further introduces three other losses added together, representing the KD on frame-level, sequence-level, and context-level. This approach serves to distill information from a pretrained audio speech recognizer to the visual model, on the encoder and decoder levels. On the other hand, \cite{c8} use state-of-the-art ASR acoustic model by adapting it for their lipreading model incorporating only frame-wise KD.

%------------------------------------------------------------------------

\section{Method}

Inspired by \cite{withoutpain} and \cite{c14}, we propose an architecture divided into the visual model and the audio model, as shown in Figure \ref{pipeline}. In the visual model, we follow the architecture proposed by \cite{withoutpain}; for the audio model, our proposed model follows the same architecture of the visual model, since it is proven to achieve better results than having different architectures for the visual and audio models \cite{c14}.
\subsection{Data processing}
\paragraph{Word Boundary.} Word boundary is implemented by introducing a binary indicator at each time step and then concatenating this indicator with the original visual frames in the front-end network. Proven to improve the results of word-based lipreading models on the LRW dataset \cite{withoutpain}, this method helps the network identify the video segment that it should focus on (in the LRW dataset, the spoken word is at the middle of the video frame).

\paragraph{Visual frames cropping and gray-scale conversion.} To narrow down the region of interest, the model receives input in the form of videos, which is resized to 96x96 and randomly cropped to 88x88 to make use of the different contextual visual features. The input is then converted to gray-scale, to remove the color as a factor when training since its does not add much extra information \cite{withoutpain,c4}.

\subsection{Data augmentation and regularization} 
In addition to the data processing tasks, several techniques are applied to help the model generalize better and avoid over-fitting, since the LRW dataset has a tendency to overfit on the training data \cite{withoutpain}.

\paragraph{Mixup.} Proven to enhance the performance of LRW visual recognizers as demonstrated in \cite{withoutpain}, mixup is implemented by selecting two samples to generate a new sample using a weighted linear interpolation in both the ground truth and the loss function as a form of augmentation. For the reader's convenience we cite the equations below.
\begin{equation}
    \hat{x} = \lambda x_{A} + (1 - \lambda) x_{B}, \hat{y} = \lambda y_{A} + (1 - \lambda) y_{B},
\end{equation} where $\lambda$ is an arbitrary value between 0 and 1.

\paragraph{Label smoothing.} As a form of regularization, and following what was proposed in \cite{withoutpain}, label smoothing is meant to modify the construction of $q\textsubscript{i}$ in the cross entropy loss function instead of the following: 
\begin{equation}
    L = -\sum_{i=1}^{N}q_{i}\log(p_{i})\left\{\begin{matrix}
q_{i}=0, y\neq i\\ 
q_{i}=1, y=i
\end{matrix}\right.
\end{equation}
to be equal to:
\begin{equation}
    q_{i} = \left\{\begin{matrix}
\frac{\epsilon}{N},  y\neq i\\ 
\\
1-\frac{N-1}{N}\epsilon,  y=i
\end{matrix}\right. 
\end{equation}
where $\epsilon$ is a small constant. In our implementation, we used the same value used by \cite{withoutpain}, which is equal to 0.1.
\subsection{Model architecture}

\paragraph{Frontend.} In the frontend, we use Squeeze and Excitation (SE) ResNet blocks since they outperform traditional ResNets due to their use of attention mechanisms to model channel-wise relationships. They also enhance the representation ability of the module through the network \cite{withoutpain}. The ResNet blocks are followed by global average pooling and dropout.

\paragraph{Backend.} We incorporate Knowledge Distillation (KD) to make use of the dual nature of the audio-visual data. Initially, we use a 3-layered bidirectional GRUs as the main recurrent building block, with dropout, average pooling, followed by a fully connected layer, then a softmax layer for classification. The final softmax layer outputs the prediction of the most probable word pronounced in the video input, where the output dimension is equal to the total number of word classes. In LRW, the number of classes is 500 words.

\section{Knowledge distillation incorporation}

\begin{figure} [t]
\begin{center}
\includegraphics[width=8cm]{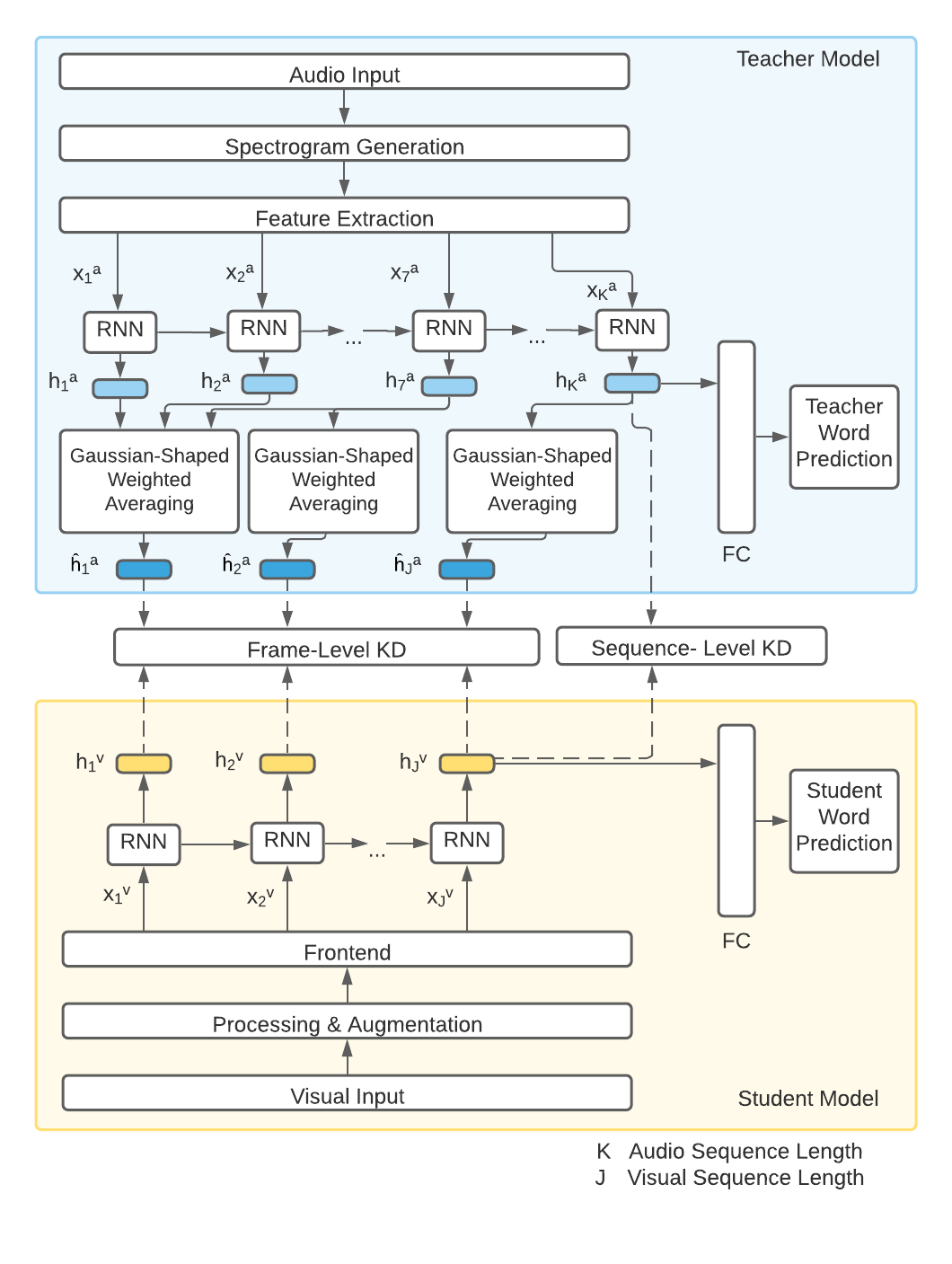}
\end{center}
\caption{Knowledge Distillation Incorporation}
\label{kd}
\end{figure}

\subsection{Teacher model} 
To implement knowledge distillation, we first need to build the pretrained teacher model, namely the audio recognizer. The following sections serve to illustrate the different steps for building the teacher model. 

Audio signal extraction and data preparation is first required for the audio recognizer to train. We extract the .mp3 files from the videos that exist in the dataset, then generate the spectrograms, using a sampling rate of 16 KHz. Then, Augmentation and preprocessing is applied to increase the diversity of the dataset and improve the accuracy of the model. The following techniques are applied: 

\begin{itemize}
  \item \textbf{Spectrogram augmentation:} Inspired by \cite{Park_2019}, we applied spectrogram augmentation due to its efficiency and simplicity as compared to standard augmentation techniques such as changing the pitch, speed, injecting noise, and adding reverb to the audio data. We implemented frequency and time masking as augmentation methods to improve generalization.
  \item \textbf{Word isolation:} To get rid of the irrelevant information at the beginning and at the end of the audio samples, we apply word isolation by AND-ing the generated waveform with zeros outside the word boundaries. This process helps maintain the size of the audio data samples needed as input to the audio model.
\end{itemize}

The architecture of the teacher model starts with three residual convolutional blocks to learn the relevant audio features. They are then followed by a linear layer and BiGRU layer with the same dimensions as the one presented in the visual model to learn the sequential data. Lastly, a fully connected linear layer followed by a softmax layer for the classification. 

The reason behind the selection of this architecture among many from the literature is that it’s similar to one of the most popular end-to-end models, namely Deep Speech 2 by Baidu \cite{deepspeech2}. Additionally, both visual and audio models have the same architecture in the backend, including the same number of BiGRU layers, batches, and hidden size. This approach is based on the experiments done by \cite{c14} concluding that the KD has the best effect when both models have the same architecture, even if the audio recognizer performance on its own is not the best. 

\subsection{Student model} 
Given the nature of the LRW dataset, speech recognition becomes more of a classification problem rather than having a sequence of words as an output. Unlike \cite{c14}, our model only makes use of the frame-level and sequence-level KD by distilling knowledge from the pretrained audio recognizer.
\begin{equation}
    L_{total} = L_{base} + \lambda_{1}L_{KD_{1}} + \lambda_{2}L_{KD_{2}}, 
\end{equation} where L\textsubscript{KD\textsubscript{1}} and L\textsubscript{KD\textsubscript{2}} are sequence-level and frame-level KD losses respectively, and $\lambda_{1}$ and $\lambda_{2}$ are the corresponding balance weights, having the values of 2 and 10, respectively \cite{c14}.

\paragraph{Sequence-level KD.} The sequence vectors, which are the output of the encoder, contain the semantic information of the input sequence, so we calculate the Mean Square Error (MSE) between the visual sequence vector and the audio sequence vector. Since both sequence vectors are of the same size, no further linear or affine transformation is needed, as shown in the following equation where  $s\textsuperscript{a}$ and $s\textsuperscript{v}$ are  the  audio  and  visual  sequence  vectors, respectively.

\begin{equation}
L_{KD_{1}} = \left \| s^{a}-s^{v} \right \|_{2}^{2}
\end{equation}

\paragraph{Frame-level KD.} Due to the different sequence lengths between the audio features and visual features, which are 139 and 29 respectively, every video frame corresponds to about 4.8 audio frames. To calculate the mean square error between the hidden states, we need to first learn the correspondence between the audio and visual features. This correspondence is implemented using a Gaussian-shaped weighted average. Unlike \cite{c14}, where the correspondence is learnt using a technique that is similar to the attention mechanism, the Gaussian-shaped weighted average offers a simpler and more direct way to map the audio hidden states to the visual hidden states using a normal distribution for the frames weights while keeping in mind the efficiency of the model training time.  

We use a sliding window of 7 points, where the mean is at the corresponding middle audio feature. This alignment gives more weight to the frames in the middle and less weights as we deviated from the middle audio frame, since the middle audio frames serve to be more correlated to the visual frame. These 7 weights are then padded with zeros across the samples outside the specified window. Applying this sliding window averaging across the audio hidden states is then used to generate $\tilde{h}$ that is used to calculate the third loss component as shown in the following equation, where J is the number of visual frames.
\begin{equation}
    L_{KD_{2}} = \frac{1}{J}\sum_{j=1}^{J}\left \| h_{j}^{v}-\tilde{h}_{j}^{a} \right \|_{2}^{2}
\end{equation}

\section{Experiments}
{

\subsection{KD Model Experimental setup}

The model is trained on the Lip Reading in the Wild (LRW) dataset; a word-based dataset comprised of 1000 utterances of 500 classes of English words, as spoken in various BBC programs. A single utterance is comprised of 29 frames (1.16 seconds), with the desired word located at the center. We trained the model on a dual-GPU machine, with each GPU assigned a batch size of 8 due to limited available computational resources. To avoid over-fitting, a cosine learning rate is utilized with the initial learning rate set to 3e-4. The estimated time per epoch, is approximately 2.7 hours.  

}

\subsection{Audio model results}

\begin{table}
\begin{center}
\begin{tabular}{|l|c|}
\hline
Method & Top-1 Acc. (\%) \\
\hline\hline
Basic Model & 86.23 \\
Basic Model + WI + partial SA & 97.56 \\
\hline
\end{tabular}
\end{center}
\caption{Audio Model Results}
\label{audio}
\end{table}

The results in table \ref{audio} represent experiments for the teacher model. We experimented the effect of incorporating Word Isolation technique (WI) and Spectrogram Augmentation (SA). We applied word isolation in the data preprocessing, to not be distracted by the additional words or letters at the beginning and at the end of each word, as previously mentioned. Also, instead of applying the spectrogram augmentation on the whole dataset, we specified the training data to be augmented during preprocessing. Additionally, we decreased the initial learning rate from 2e-4 to 1e-4. These modifications increased the accuracy of the teacher model by 5.44\%.

\subsection{Visual model results}
We re-trained the final model proposed by \cite{withoutpain} as our baseline model. Our results when re-training, achieved an accuracy of 87.82\% which is 0.58\% lower than the reported accuracy, 88.4\%. The difference in performance might be due to the changes in the hyper-parameters, namely the batch-size, learning-rate and number of GPUs. It can also be due to the fact that in \cite{withoutpain}, Dalu Feng \etal did not specify the seed used in order to replicate their results.

\begin{table}
\begin{center}
\begin{tabular}{|l|c|}
\hline
Method & Top-1 Acc. (\%) \\
\hline\hline
Baseline & 87.82 \\
Baseline + KD\textsubscript{1} & 88.25 \\
Baseline + KD\textsubscript{1} + KD\textsubscript{2} & \textbf{88.64} \\
\hline
\end{tabular}
\end{center}
\caption{Effect of Different Levels of KD}
\label{kd}
\end{table}

The accuracy scores reported in table \ref{kd} demonstrate the effect of different levels of knowledge distillation on the performance of the model. It is noticeable that the addition of the KD losses to the main loss function enhances the performance as compared to the baseline model by 0.4\% and 0.84\% for the sequence-level KD\textsubscript{1} and the sequence-level KD\textsubscript{1}+frame-level KD\textsubscript{2}, respectively. Additionally, the value of the STD used for the Gaussian-shaped average affects the performance of the model. We experimented with different STD values equal 3 and 2. They achieved Top-1 accuracy scores of 88.64\% and 88.14\% respectively.

\section{Conclusions and future work}
This work sets a new benchmark on the LRW dataset using a variant of knowledge distillation, outperforming state-of-the-art results by a noticeable margin. The proposed architecture serves to leverage both audio and visual data to enhance the performance of the model, highlighting the importance of "listening" to the lips before "reading" them. While \cite{c14} initially introduced this technique for the sentence-based models, this work presents the first word-level lipreading model utilizing audio-visual data. Future works can build on the proposed model by leveraging different types of unlabeled data, such as YouTube videos, to further enhance the accuracy.

Additionally, the proposed model may be tweaked further in order to increase it's efficiency by lowering its training time. In the case of more powerful machines being available, the batch size and number of GPUs can be  increased accordingly. Additionally, further enhancements in the fronend and preprocessing stage can be implemented by introducing new augmentation techniques, such as Attentative CutMix, or by integrating Spatio-Temporal Attention in the frontend to better extract facial features. Finally, the model will be publicly release to support the reproducibility of the submitted work.

\small
\bibliographystyle{IEEEtranS}
\bibliography{egbib}

\end{document}